\documentclass{nature}
\usepackage{aas_macros,color}
\usepackage{graphicx}
\def\src {SGR\,0418+5729}

\title{A variable absorption feature in the X-ray spectrum of a magnetar}

\author{Andrea Tiengo$^{1,2,3}$, Paolo Esposito$^2$, Sandro Mereghetti$^2$, Roberto Turolla$^{4,5}$, Luciano Nobili$^{4}$, Fabio Gastaldello$^2$, Diego G{\"o}tz$^6$, \mbox{Gian Luca} Israel$^7$, Nanda Rea$^8$, Luigi Stella$^7$, Silvia Zane$^5$ \& Giovanni~F.~Bignami$^{1,2}$}

\begin{document}

\maketitle

\begin{affiliations}
 \item Istituto Universitario di Studi Superiori, piazza della Vittoria 15, I-27100 Pavia, Italy.
 \item Istituto di Astrofisica Spaziale e Fisica Cosmica Milano, INAF, via Bassini 15, I-20133 Milano, Italy.
 \item   Istituto Nazionale di Fisica Nucleare, Sezione di Pavia, via Bassi 6, I-27100 Pavia, Italy.
 \item Dipartimento di Fisica e Astronomia, Universit\`a di Padova, via Marzolo 8, I-35131 Padova, Italy.
 \item Mullard Space Science Laboratory, University College London, Holmbury St.~Mary, Dorking, Surrey RH5 6NT, UK.
\item AIM CEA/Irfu/Service d'Astrophysique, Orme des Merisiers, F-91191 Gif-sur-Yvette, France.
 \item  Osservatorio Astronomico di Roma, INAF, via Frascati 33, I-00040 Monteporzio Catone, Italy.
\item Institut de Ci\`encies de l'Espai (IEEC--CSIC), Campus UAB,  Torre C5, 2a planta, E-08193 Barcelona, Spain.
\end{affiliations}

\begin{abstract} Soft-$\gamma$-ray repeaters (SGRs) and anomalous X-ray pulsars (AXPs) are slowly rotating, isolated neutron stars that sporadically undergo episodes of long-term flux enhancement (outbursts) generally accompanied by the emission of short bursts of hard X-rays\cite{mereghetti08,rea11}. This behaviour can be understood in the magnetar
model\cite{thompson95,thompson96,tlk02}, according to which these sources are mainly powered by their own magnetic energy.
This is supported by the fact that the magnetic fields inferred from several observed properties\cite{kouveliotou98,thompson01,vietri07} of AXPs and SGRs are greater than -- or at the high end of the range of -- those of radio pulsars.
In the peculiar case of SGR 0418+5729, a weak dipole magnetic moment is derived from its timing parameters\cite{rea13}, whereas a strong field has been proposed to
reside in the stellar interior\cite{rea10,turolla11} and in multipole components on the surface\cite{guver11}. Here we show that
the X-ray spectrum of \src\ has an absorption line, the
properties of which depend  strongly on the star's rotational phase. This line is interpreted as a proton cyclotron feature and its energy implies a magnetic field ranging from 2$\times$10$^{14}$ gauss to more than 10$^{15}$ gauss.
\end{abstract}

On 2009 June 5 two short bursts of hard X-rays, detected by Fermi and other satellites, revealed the previously unknown source
\src\cite{vanderhorst10}. Subsequent observations with the Rossi X-ray Timing Explorer (RXTE), Swift, Chandra and X-ray Multi-mirror
Mission (XMM) Newton satellites found the new SGR to be an X-ray pulsar with a period of $\sim$9.1 s and a luminosity of $\sim$$1.6\times10^{34}$ erg
s$^{-1}$ (in the 0.5--10 keV band and for a distance of 2 kpc)\cite{vanderhorst10,esposito10}. During the three years after
the onset of the outburst, the spectrum softened and the luminosity declined by three orders of magnitude, but remained still too high
to be powered by rotational energy\cite{esposito10,rea10,rea13}.
The measured spin-down rate of $4\times10^{-15}$ s s$^{-1}$ translates (under the assumption of  rotating magnetic dipole \emph{in vacuo}) into a magnetic field $B = 6\times10^{12}$~G at the magnetic equator\cite{rea13}, 
a value well in the range of  normal radio pulsars.
However, the presence of high-order multipolar field components of $10^{14}$~G close to the surface has been invoked to interpret the spectrum of the source in the framework of atmosphere models\cite{guver11}. In any case, a strong crustal magnetic field ($>10^{14}$~G) seems to be required to explain the overall properties of \src\ within the magnetar model\cite{turolla11,rea13}. 

Hints of the presence of an absorption feature at 2 keV in the spectrum of \src\ were found in the phase-resolved analysis of data (with
relatively low-count statistics) from the Swift X-ray Telescope (XRT) taken during 2009 July 12--16\cite{esposito10}. Thanks to the
large collecting area and good spectral resolution of the European Photon Imaging Camera (EPIC), we were able to perform a more detailed investigation using data collected by XMM-Newton during a 67-ks long observation performed on 2009 August 12, when the source
flux was still high ($5\times10^{-12}$ erg cm$^{-2}$ s$^{-1}$ in the 2--10 keV band).

To examine
the spectral variations as a function of the star's rotational phase without making assumptions about the X-ray spectral energy distribution of  \src, we produced a
phase--energy image
by binning the EPIC source counts
into energy and rotational phase channels and then normalising to the phase-averaged energy spectrum and pulse profile.
The normalised phase--energy image (Fig.1)   shows a prominent V-shaped feature in the phase interval $\sim$0.1--0.3.
This is produced by a lack of counts in a narrow energy range with respect to nearby energy channels, that is, an absorption
feature at a phase-dependent energy.
The regular shape of the feature in the phase--energy plane as well as its presence in the three independent EPIC detectors (see Supplementary Fig.\,5) exclude the possibility that it results from statistical fluctuations in the number of counts or
from an instrumental effect.
Another absorption feature is visible at low energies at phase $\sim$0.5--0.6.

We extracted from the EPIC data the phase-averaged spectrum of \src, as well as the spectra from 50 phase intervals of width 0.02 rotational cycles, as described in the Supplementary Information.  The phase-averaged spectrum can be adequately fit by either
a two-blackbody model ($\chi^2_{\nu}=1.198$ for 196 degrees of freedom, d.f.) or a blackbody plus power-law model ($\chi^2_{\nu}=1.105$ for 196 d.f.), corrected for interstellar absorption
(see refs 11 and 12 for other models that can fit the
spectrum).

The 15 spectra extracted from the phase intervals 0.1--0.3
and 0.5--0.6, unlike those of the remaining phases, cannot be fitted
by a renormalisation of the phase-averaged best-fit model,  which
gives in most  cases null hypothesis probabilities in the range $10^{-4}$ -- $10^{-9}$ (see Supplementary Fig.\,4). They are instead well fitted (null hypothesis
probability $>$0.03) by the addition of a narrow absorption line
component, which can be equally well modelled with a Gaussian
profile or a cyclotron absorption line model\cite{makishima90}
(the improvement obtained by adding a cyclotron component in the
phase intervals 0.1--0.3   and 0.5--0.6 can be seen in Supplementary Fig.\,4).
 The best-fit line parameters as a function of phase are shown in Fig.\,2 and an example of phase-resolved spectrum is displayed in Fig.\,3.

We searched for the phase-dependent absorption feature in all the available X-ray observations of \src\ and  found that it was present in the phase interval 0--0.3, and up to higher energies than in XMM-Newton,  in RXTE data taken during the first two months of the outburst (see Supplementary Fig.\,6).

Absorption features have been observed in the X-ray spectra of various classes of neutron
stars\cite{truemper78,heindl04,turolla09,haberl06,vankerkwijk07,bignami03,gotthelf13,kdm12} and interpreted as being due to
either  cyclotron absorption  (by electrons or protons) or bound--bound atomic transitions. However, variations in the line energy as a function of the
rotational phase as large as in \src\  (by a factor $\sim 5$ in one-tenth of a cycle) have not been seen in any source.

In a neutron star atmosphere, different atomic transitions might be responsible for a phase-variable absorption feature if temperature, elemental abundance or magnetic field vary strongly on the surface.
The line energies observed in \src\ ($\sim$1--5 keV) rule out transitions in magnetised H and He, which occur below $\sim$1~keV\cite{potekhin98,medin08}. On the other hand, the absorption spectra of heavier elements are much more complex (see, for example, ref.\,26 for C, O and Ne) and some lines could occur at high-enough energies.
However, to explain the phase resolved spectra of \src, the physical conditions of a heavy-element atmosphere are forced to vary in such a way that a 
single transition should dominate the opacity at each of the phases where 
the absorption line is detected.

A more straightforward explanation for the line variability can instead be given if the feature is due to cyclotron resonant scattering. The cyclotron energy (in keV) for a particle of charge $e$ and mass $m$ in magnetic field $B$ (in gauss) is given by 
\begin{displaymath} E_B\approx
\frac{11.6}{1+z}\left(\frac{m_e}{m}\right)\left(\frac{B}{10^{12}}\right) \end{displaymath} where
$(1+z)^{-1}= (1-2GM_{\mathrm{NS}}/(Rc^2))^{1/2}$ (which is $\sim$0.8 at the star surface
for typical neutron star mass and radius $M_{\mathrm{NS}}=1.4\,M_\odot$ and $R_{\mathrm{NS}}=12$ km, respectively)  accounts for the gravitational redshift at distance $R$ from the neutron star centre, $m_e$ is the mass of the electron,
and $c$ is the velocity of light.
In this case, 
the phase variability of the feature energy would simply be due to the different fields experienced by the charged particles interacting with the photons directed towards us as the neutron star rotates.

If the absorbers/scatterers are electrons hovering near the star surface, the expected line energy is
$\sim$70~keV for the dipole field at
the equator of \src\ ($B= 6\times 10^{12}$~G); this line energy is more than 10
times higher than that observed. A possible way to explain this large discrepancy might be to assume that the electrons producing the
line are located higher up in the magnetosphere in a dipolar geometry, where the magnetic field is
smaller ($R\approx 3R_{\mathrm{NS}}$ to have $E_B\approx 2\ {\rm keV}$).
Moreover, such an electron population should also be nearly mono-energetic, or subrelativistic, in
order to prevent Compton scattering from washing out the feature, which would require a mechanism to maintain slowly moving  electrons confined in a small volume high in the magnetosphere.

If the particles responsible for the cyclotron scattering are protons, the energy range of the \src\ spectral feature requires a magnetic field $>2\times 10^{14}$ G (it would be even larger for heavier ions). In the framework of the magnetar model, the unprecedented phase-variability of the line energy can be explained by the
complex topology of the magnetar magnetospheres, in which global
and/or localised twists play an important part\cite{tlk02}. 
This is particularly true for \src, which has a weak dipolar
component, as testified by the small spin-down value, whereas 
a much stronger internal magnetic field has been advocated
to explain its X-ray luminosity and burst activity\cite{rea10,turolla11}.
Furthermore, the presence of small-scale, strong, multipolar components of the surface field 
has been inferred by fitting its phase-averaged X-ray spectrum with models of magnetized neutron star atmospheres\cite{guver11}.

In this context, the observed line variability might be due to the presence of either strong magnetic field gradients along the surface or vertical structures (with a spatially dependent field) emerging from the surface.
To work out how the dynamic magnetosphere of a magnetar should look, an analogy with the solar corona in the proximity of sunspots has been proposed (see, for example, ref.\,27). In particular,
localised, baryon-rich, magnetic structures, in the form of rising flux tubes, or `prominences', produced by magnetic reconnection or the emergence of the internal field near a crustal fault, have been proposed to explain some of the observed properties of the giant flare emitted in 2004 by SGR\,1806--20\cite{gelfand05,masada10}. If a similar scenario, albeit on a reduced scale, occurred during the outburst of \src, a spectral feature might arise as thermal photons from the hot spot (a small hot region on the neutron star surface, responsible for most of the X-ray emission, which could be itself related to the prominence) cross the plasma threading the magnetic loop. A proton density $\approx$$10^{17}$ cm$^{-3}$ is needed to produce a resonant scattering depth of order unity\cite{tlk02}. Protons, being heavy, do not rise much above the surface and move subrelativistically\cite{tlk02}, so resonant scattering in the prominence is likely to produce a narrow feature instead of an extended tail. As the star rotates, photons emitted in different directions
pass through portions of the prominence with different magnetic field, density and size, giving rise to the observed variations of the line centroid and width.
A simple quantitative
model based on this picture is presented in Supplementary Information. Results,
obtained with a geometry consistent with the constraints derived
from the X-ray pulsed fraction of \src, are in good agreement
with the observed variations of the feature with phase (Fig.\,1).

\begin{addendum}
\item
We thank G. Goggi and C. Paizis for useful discussion.
This research is based on data and software provided by the  the ESA XMM-Newton Science Archive (XSA) and the NASA/GSFC High Energy Astrophysics Science Archive Research Center (HEASARC). The Authors acknowledge partial funding from INAF through a PRIN 2010 grant and ASI through contract I/032/10/0.
\item[Author Information] 
The authors declare that they have no competing financial interests. 
Correspondence and requests for materials
should be addressed to A.T.~(email: andrea.tiengo@iusspavia.it).
\end{addendum}

\clearpage

\bibliography{biblio}

\clearpage

\begin{figure}
\includegraphics[width=\textwidth]{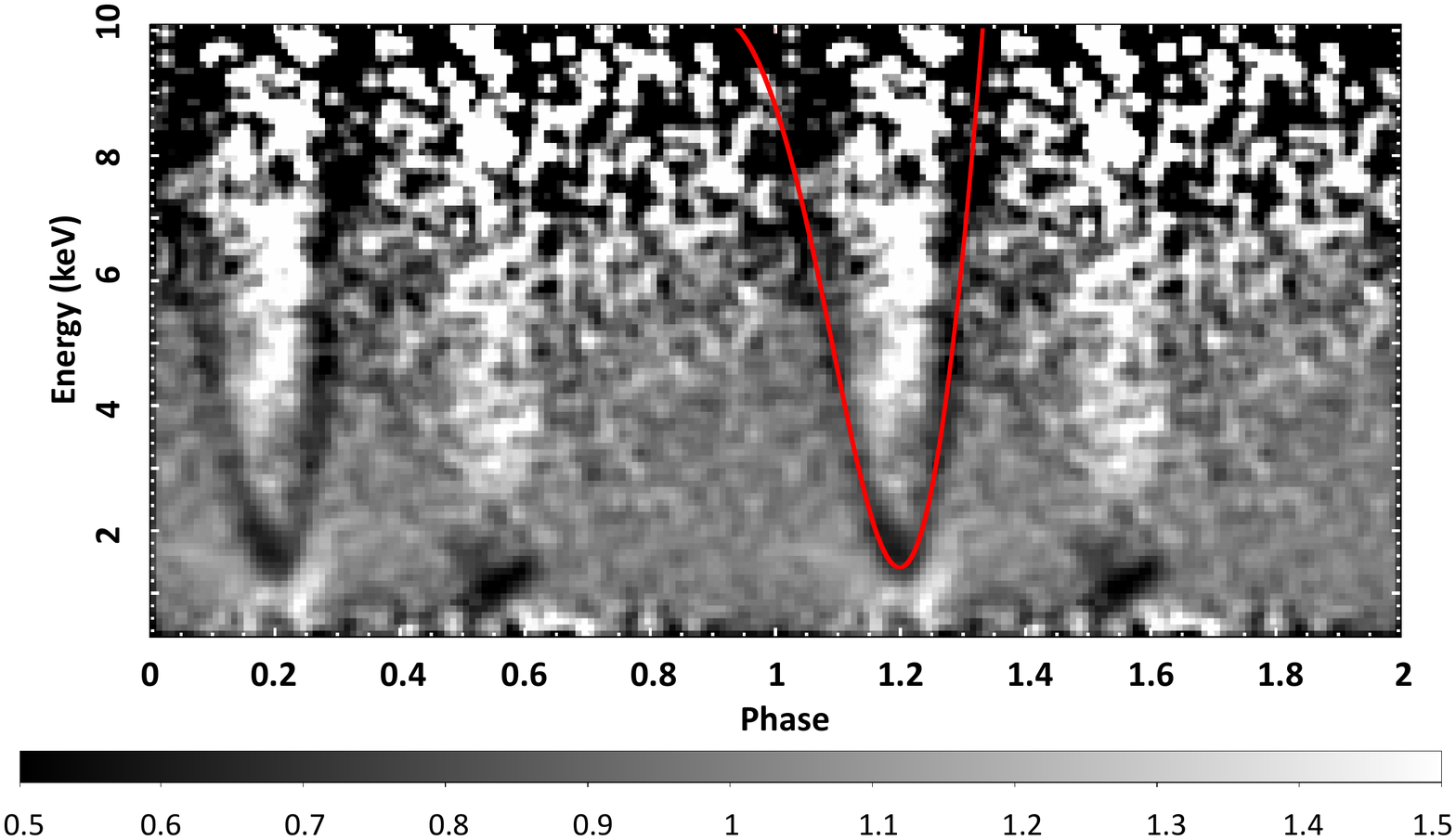}
\caption{\textbf{$|$ Phase-dependent spectral feature in the EPIC data of \src.}
Normalised energy versus phase image obtained by binning the EPIC source counts into 100 phase bins and 100-eV-wide energy channels and dividing these values first by the average number of counts in the same energy bin (corresponding to the phase-averaged energy spectrum) and then by the relative 0.3--10 keV count rate in the same phase interval (corresponding to the pulse profile normalised to the average count rate).
The red line shows (for only one of the two displayed cycles) the results of a simple proton cyclotron model consisting of a baryon-loaded plasma loop emerging from the surface of a magnetar and intercepting the X-ray radiation from a small hotspot (see Supplementary Fig.\,7 and Supplementary Table\,1).
 }
\end{figure}

\begin{figure}
\includegraphics[width=0.75\textwidth]{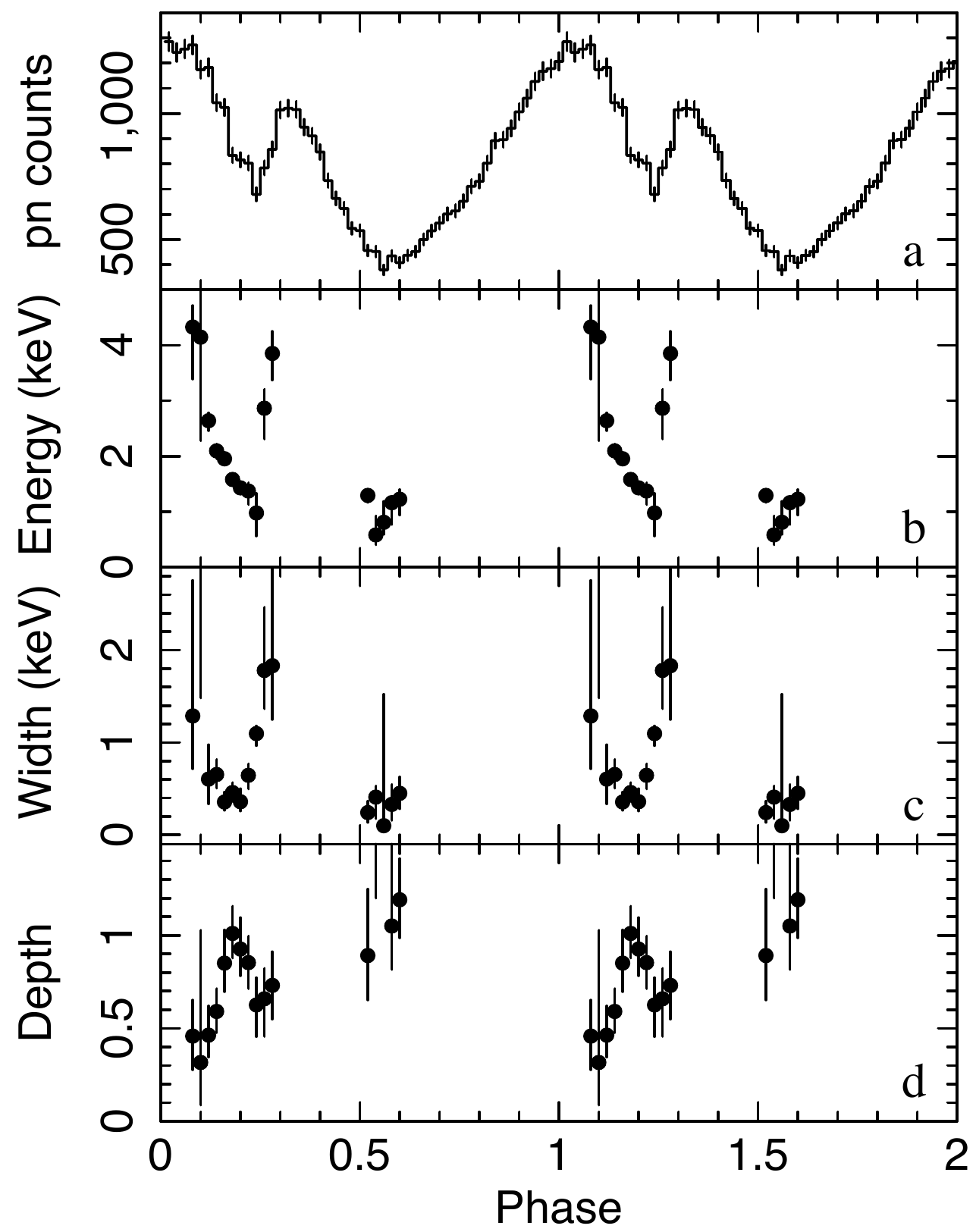}
\caption{\textbf{$|$ Results of the phase-resolved spectroscopy of \src.}
\textbf{a}, Pulse profile obtained by folding the 0.3--10 keV EPIC pn light curve at the neutron star spin period  $P=9.07838827$ s. The data points are the number of counts in
each phase-dependent spectrum. \textbf{b-d}, Line energy ($E_{\rm c}$; \textbf{b}),  width ($W$; \textbf{c}) and depth ($D$; \textbf{d}) of the cyclotron feature as a function of the spin phase. The model consists of a blackbody plus a power law and an absorption line, modified for the interstellar absorption (see Supplementary Information).  For the line we used the cyclotron absorption model from ref.\,15: $F(E)=\exp\Big(-D\frac{(W E/E_{\rm c})^2}{(E-E_{\rm c})^2 +W^2}\Big)$.  The interstellar absorption, temperature, photon index and relative normalisations of the two components were   fixed to the best-fit values of the phase-integrated spectrum: $N_{\rm{H}}= (9.6\pm0.5)\times 10^{21}$ cm$^{-2}$, $kT=913\pm8$ eV, $\Gamma=2.8\pm0.2$, $(R_{\rm BB}/d)^2=0.81\pm0.03$ km$^2$ kpc$^{-2}$ and $K_{\rm PL}=(1.5\pm0.2)\times10^{-3}$ photons cm$^{-2}$ s$^{-1}$ keV$^{-1}$ at 1 keV. Vertical error bars, 1 s.d.}
\end{figure}

\begin{figure}
\includegraphics[width=0.7\textwidth]{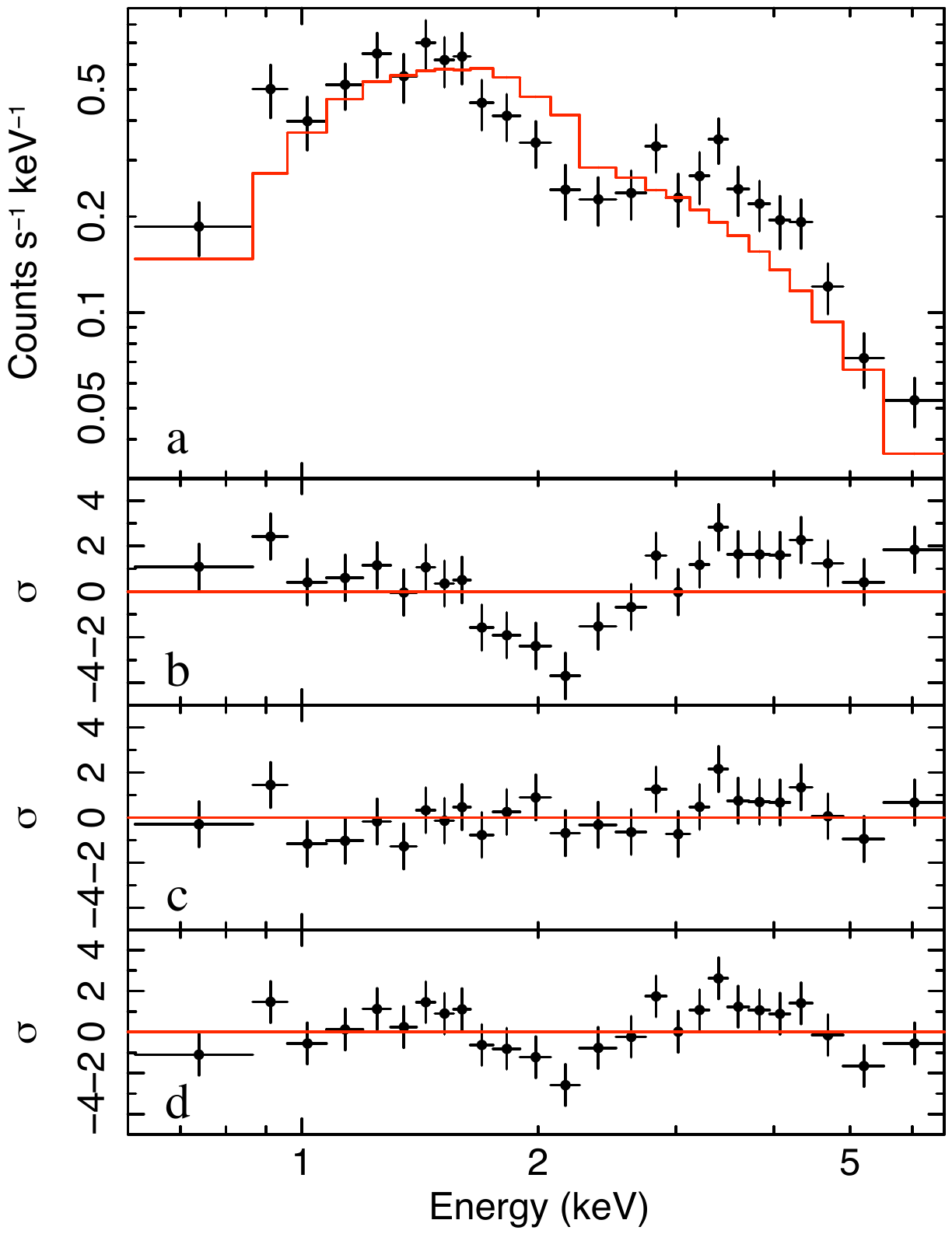}
\caption{\textbf{$|$  Example of a phase-resolved EPIC pn spectrum and its residuals with respect to different models.}
\textbf{a}, Spectrum  from the phase interval 0.15--0.17 (black dots) and best-fit model of the phase-averaged spectrum, rescaled with a free normalisation factor (red line).  \textbf{b}, Residuals with respect to this model ($\chi^2_{\nu}=2.75$ for 25  d.f.); \textbf{c},  residuals after the addition of an absorption line (\emph{cyclabs} model in \emph{XSPEC}, with parameters as in Fig.\,2; $\chi^2_{\nu}=0.94$ for  22 d.f.); \textbf{d}, residuals with respect to an absorbed blackbody plus power law model with free temperature, photon index and normalisations ($kT=1.11\pm0.06$ keV and $\Gamma=3.8\pm0.4$; $\chi^2_{\nu}=1.75$ for 22  d.f.). This is one of the models (with the same number of free parameters) that we also explored to fit the phased-resolved spectra. 
In this case, we obtained fits of comparable quality to those with the line model at most phases, but worse fits in the phase interval 0.11--0.21. A joint fit to these five spectra gave an unacceptable $\chi^2_{\nu}$ of 1.56 for 116 d.f., to be compared with $\chi^2_{\nu}$ of 1.08 for the absorption line model (same number of d.f.). Horizontal error bars indicate the energy channel width; vertical error bars, 1 s.d.; residuals $\sigma$ are in units of standard deviations.}
\end{figure}

\clearpage

\section{XMM-Newton data analysis}

The data have been processed with version 12 of the
\emph{Scientific Analysis Software} (\emph{SAS}) and we used the
most recent (2012 October) calibration files available for the
EPIC instrument. EPIC consists of two MOS$^{30}$ and one
pn$^{31}$ CCD cameras sensitive to photons with energy
between 0.2 and 10 keV. During the 2009 observation (see also
Refs.\,32,\,33), the two MOS and the pn cameras
were set in Small Window mode (time resolution of 0.3 s and 5.7
ms, respectively); all detectors were operated with the thin
optical blocking filter. Periods in which the particle background
was unusually high because of soft proton flares were excluded
using an intensity filter. This reduced the net exposure time to
30.7 ks, 50.0 ks and 50.7 ks for pn, MOS1 and MOS2, respectively.
Again according to standard procedure, photon event grades higher than 12
for the MOS cameras and 4 for the pn were filtered out. Photon
arrival times were converted to the Solar System barycentre
reference frame, by using the coordinates$^{34}$ $\rm
R.A. = 04^h 18^m 33.87^s$, $\rm Decl.=+57^\circ 32' 22.91''$
(J2000) and the spin phases were computed with the timing
parameters of Ref.\,32 ($P=9.07838827$, Epoch 54993 MJD,
valid over the range MJD 54993--55463). The phase--energy images
extracted for each of the three EPIC cameras are shown in Fig.\,5.
The V-shaped feature indicating the presence of a phase-dependent
absorption line is present in the three independent datasets.\\

We also analysed in the same way the three XMM-Newton observations
of \linebreak \src\ performed from September 2010 to August 2012 (the
observation settings and the source flux are reported in
Ref.\,35). However, due to the lower flux, they  were
not sensitive enough to detect the absorption feature, even if it
were still present.\\

For the source spectra we used the pn counts extracted from a
circular region with radius of 35$''$; the background spectra were
extracted from source-free regions on the same chip as the target.
The ancillary response files and the spectral redistribution
matrices were generated with the \emph{SAS} tasks \emph{arfgen}
and \emph{rmfgen}, respectively. For the spectral analysis,
performed in the 0.3--10 keV energy range, we used the
\emph{XSPEC} fitting package version 12.4$^{36}$. The
abundances adopted were those of Ref.\,37 and for the
photoelectric absorption we used the cross-sections from
Ref.\,38.

\section{RXTE and Swift data analysis}

 The RXTE and Swift observations used in this research have been
 already presented in Ref.\,39, to which we refer
for more information. All data were reprocessed and analysed with
version 6.12 of the \emph{HEAsoft} package and the \emph{CALDB}
calibration data base available in 2013 February. Apart from this,
the Swift data were reduced exactly as described in
Ref.\,39.\\

For each RXTE/PCA data-set, we ran the \emph{xenon2fits} script to
combine the GoodXenon files into science event tables with 256
energy bins. We used only data from the Proportional Counter Unit
PCU-2, since it is the best-calibrated unit of the PCA instrument,
and selected photons from the layer 1 of the detector. We
corrected in the event tables the photon arrival times to the
Solar System barycentre using the script \emph{fxbary}.\\

Fig.\,6 shows the phase--energy images (rescaled to the
phase-averaged spectrum) obtained from PCA observations of \src\
in three different epochs.
phase-dependent absorption feature is apparent in the phase
interval 0--0.3, and up to higher energies ($\sim$10 keV) than in
XMM-Newton, in the data taken in the first two months after the
onset of the outburst, and marginally visible in August 2009 due
to the low signal-to-noise ratio. The possible presence of the
line at even higher energies cannot be tested due to the very few
source photons detected at these energies.\\

The phase-resolved spectra, with 50 phase bins, were extracted
with \emph{fasebin} (which also barycentres the data) and combined
with \emph{fbadd}. Consistent phase-averaged spectra were used by
the script \emph{pcarsp} to make the 256-bin response matrices,
which were combined with \emph{addrmf}. Using the 2009
June--August spectra, we performed a similar analysis as described
for the XMM-Newton data. While the results were consistent, owing
to the limited spectral capabilities of the PCA instrument the
analysis added no new information on the
characteristics of the absorption line.\\

The phase--energy image obtained from the Swift/XRT data taken on
2009 July 12--16 is consistent with the EPIC  one. Although the
limited spectral capabilities and high instrumental background of
the  PCA instrument and the poor counting statistics of the XRT
data are not adequate for a detailed characterisation of the
feature properties, these data indicate that the absorption
feature observed with EPIC had been present in the spectrum of
\src\, for at least two months.

\section{A simple proton cyclotron resonance model}

We present here a simple model to illustrate how resonant proton
cyclotron scattering can produce a phase-variable absorption
feature in the X-ray spectrum of \src. Let us assume that thermal
emission from the star surface comes from a
small hot spot
 (for
\src\ the actual angular size is $\sim 4^\circ$ assuming a
distance of 2 kpc and a star radius $R_{\mathrm{NS}}= 10$ km; see e.g. Ref.\,33) and that an
ultra-strong, small-scale magnetic field is present above the spot
(the large scale field is a dipole with surface strength
$\sim 6\times 10^{12}$ G). The field is taken to thread a
plasma-loaded magnetic loop of radius  $r<R_{\mathrm{NS}}$,
containing (non-relativistic) protons of density $n$. For the sake of simplicity,
we  neglect the loop extent in the radial direction, while
retaining a finite transverse width, and assume that the B-field
lines are along the loop. Although many different geometries may
be envisaged, in the following the loop is assumed to have the
shape of a spherical lune (the round surface of a spherical wedge)
with the diameter on the star surface and the spot at its centre,
dihedral angle $2\beta$ and inclination $\beta_{\mathrm{c}}$ with respect to
the surface normal  (see Fig.\,7).\\

Photons of frequency  $\omega$ emitted by the spot may undergo resonant cyclotron scattering
as they traverse the loop and the optical depth is
\begin{equation}\label{tau1}
\tau_{\mathrm{\omega}}=\int n\sigma\, \mathrm{d}s=\int n\sigma_{\mathrm{T}}
(1+\cos^2\theta_{\mathrm{bk}})\delta(\omega-\omega_{\mathrm{B}})\, \mathrm{d}R\,,
\end{equation}
where the expression for the resonant cross section (e.g.
Ref.\,40) was used, $\theta_{\mathrm{bk}}$ is the angle between the
photon direction and $\mathbf B$, $R$ is the radial coordinate
counted from the spot and $\omega_{\mathrm{B}}=eB/(mc)$ is the particle cyclotron
frequency.
If $B$ is above $10^{12}$ G, photons in the $\sim
1$--10 keV range can only resonantly scatter on protons. Since we
are assuming that matter is confined in a very thin layer at $R=r$
and $\cos\theta_{\mathrm{bk}}= 0$, equation (\ref{tau1}) reduces to
\begin{equation}\label{tau2}
\tau\sim \frac{\pi^2enr}{B}\sim 0.4\left(\frac{n}{10^{17}\, {\rm cm}^{-3}}\right)
\left(\frac{r}{10^{5}\, {\rm cm}}\right) \left(\frac{B}{10^{14}\, {\rm G}}\right)^{-1}\,,
\end{equation}
so a proton density $n\approx 10^{17}\, {\rm cm}^{-3}$ is required
to make the loop thick to resonant scattering. If this occurs, the
flux $F_{\mathrm{\omega}}$ emitted by the spot will be reduced by a factor
$\exp(-\tau_{\mathrm{\omega}})$ in traversing the baryon-loaded loop. Because
only photons with $\omega=\omega_{\mathrm{B}}(r)$ do scatter, this will
produce a monochromatic absorption line at $\omega_{\mathrm{B}}(r)$. This is
clearly an effect of our approximations. The finite radial extent
of both the loop and the emitting spot will result in a broadening
of the absorption feature.\\

Thermal photons are emitted from the spot with a given angular
pattern,
but only those propagating
along the unit vector $\mathbf k$ reach the observer; $\mathbf k$
is characterized by the angle $\alpha$ it makes with the surface
normal and by the associated azimuth $\phi$. Because of general
relativistic ray bending, $\alpha$ differs from $\theta$, the
angle between the line-of-sight (LOS) and the spot normal.
We relate them using Beloborodov's approximation$^{41}$ (see also Ref.\,42)
\begin{equation}\label{gr}
\cos\alpha = \frac{R_{\mathrm{S}}}{R_{\mathrm{NS}}} +
\cos\theta\left(1-\frac{R_{\mathrm{S}}}{R_{\mathrm{NS}}}\right)\,,
\end{equation}
where $R_{\mathrm{S}}$ is the star Schwarzschild radius.\\

When the star rotates, both $\alpha$ and $\phi$ change with the
rotational phase $\gamma=2\pi t/P$, where $P$ is the star spin
period. By introducing the angles $\chi$ and $\xi$ that the LOS
and the spot normal form with the rotation axis, respectively, it is
\begin{equation}\label{trigo}
\cos\theta=\cos\xi\cos\chi-\sin\xi\sin\chi\cos\gamma\,,
\end{equation}
and
\begin{equation}\label{phi}
\cos\phi={\mathbf u}\cdot({\mathbf n}\times{\mathbf
k})/\vert{\mathbf n}\times{\mathbf
k}\vert=\frac{\sin\gamma\sin\chi}{\sqrt{1-\cos^2\theta}}\,;
\end{equation}
the latter follows from geometrical considerations (see again
Fig.\,7; note that the azimuths associated to $\alpha$ and
$\theta$ coincide) and the last equality holds for $\mathbf u$
oriented along the meridian which passes through the spot
and pointing north. The angle $\phi$ is counted
from $\mathbf u$ and changes by $2\pi$ in a cycle if it is
$\xi<\chi$ while it oscillates between a minimum and a maximum
value (which depend on $\xi$ and $\chi$) in the opposite case.\\

Because $\alpha$ and $\theta$ change during a cycle, the ray
direction ${\mathbf k}$ may or may not intersect the loop, so an
absorption feature may be present at certain phases only.
What actually happens depends on the loop geometry (here
its transverse angular width $2\beta$ and inclination
$\beta_{\mathrm{c}}$) and on the angles $\chi$ and $\xi$.
Actually,
$\chi$ and $\xi$ are not known for \src, the only
constraint coming from the measured pulsed fraction ($\sim 50\%$),
which implies that their sum is about $90^\circ$.\\

When the ray directed towards the observer crosses the
loop, the intersection will occur at different positions
according to the phase, giving rise to features at different
energies,  proportional to the local magnetic field
intensity. To obtain a phase--line energy relation to compare with
observations, we need to introduce how the magnetic field depends
on position in the loop. For simplicity, we assume that the field varies only
along the loop, linearly decreasing its intensity as its transverse width increases: 
\begin{equation}\label{bvar}
B=B_{\mathrm{max}}-f \sqrt{1 - \sin^2\alpha\cos^2(\phi-\phi_0)}\,,
\end{equation}
where $B_{\mathrm{max}}$ is the magnetic field intensity at the base of the loop (where the transverse width vanishes and the field is strongest),  $f$ is a multiplicative factor measuring how fast the field
varies along the loop and $\phi_0$ is the angle between the lune
diameter and the meridian passing through the spot. \\

Selecting convenient combinations of the parameters, our
simple analytical model can well reproduce the variation of the
line energy with phase observed in \src, including its relative
phase with respect to the maximum of the pulse profile, which
occurs at $\gamma\simeq0.05$ (see Fig.\,2) and in our model
corresponds to the phase where $\cos\alpha$ is maximum.  The red
line in Fig.\,1 shows the result obtained for a loop with
$\phi_0=90^{\circ}$ (i.e., parallel to the equator),
$\xi=20^{\circ}$,  $\chi=70^{\circ}$ (consistent with the 50\%
pulsed fraction), $\beta=30^{\circ}$ and $\beta_{\mathrm{c}}=35^{\circ}$
 (see Table\,1 for a list of all the model parameters).  
The magnetic field in the
portion of the loop swept by the line of sight varies from
$2.5\times10^{14}$ G to $5\times10^{15}$ G. Although this simple
model cannot explain with a single symmetric loop also the low
energy line observed at phase $\approx$0.55, the latter might
originate from a second loop with a smaller angular size located
on the other side of the hot spot.


\begin{figure}
\centering\includegraphics[angle=0, width=15cm]{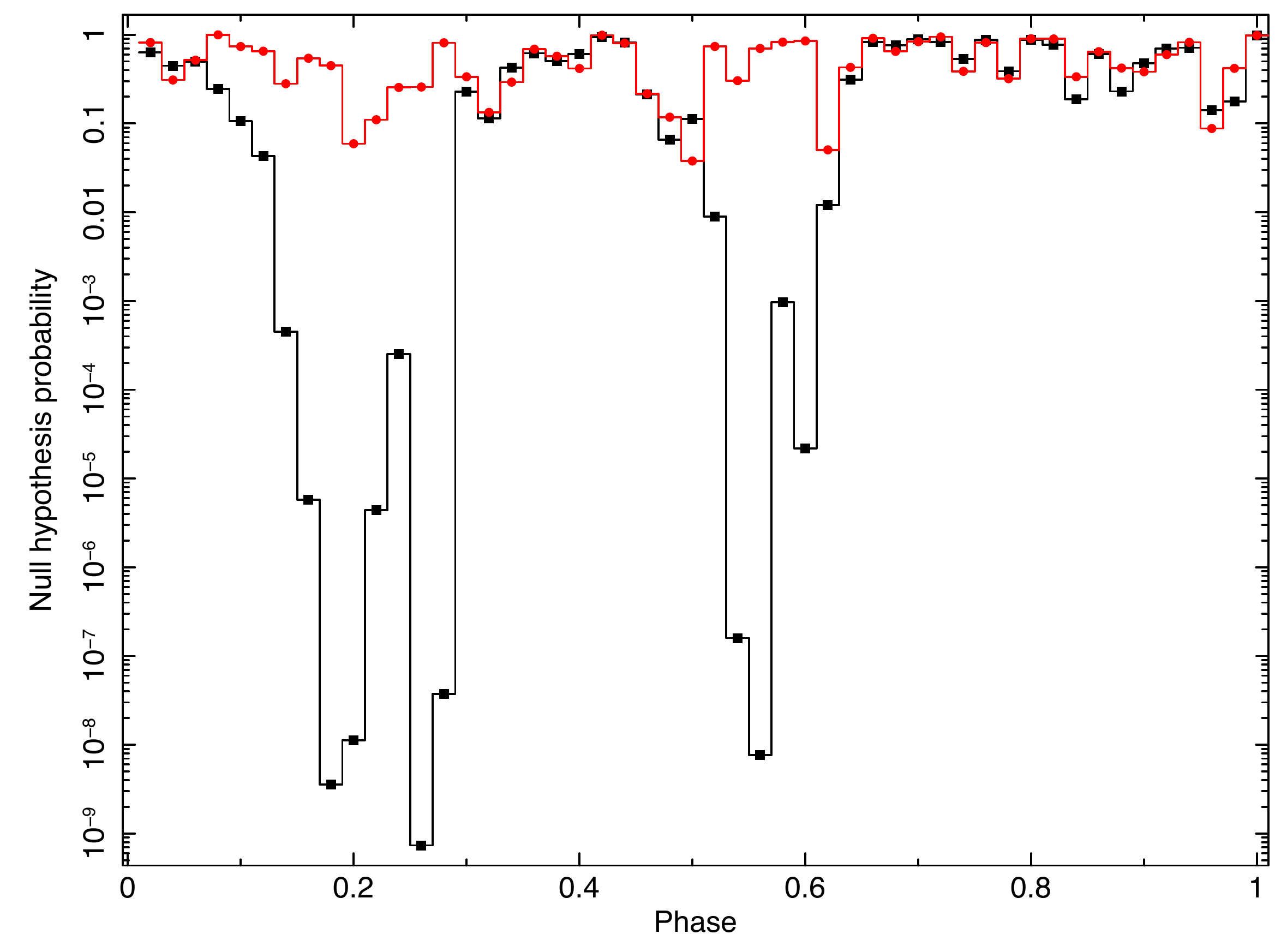}
\caption{Null hypothesis probabilities derived from the
$\chi^2$ and degrees of freedom of the fits of  the 50 EPIC pn
phase-resolved spectra with two different models: the best-fit
model of the phase-averaged spectrum (absorbed blackbody plus
power law model, with parameters fixed at the values reported in
the caption of Fig.\,2) with a free normalisation factor (black)
and this same model with the addition of an absorption line
(\emph{cyclabs} model$^{43}$ in \emph{XSPEC}; red). }
\end{figure}

\clearpage

\begin{figure}
\centering\includegraphics[angle=0, height=18cm]{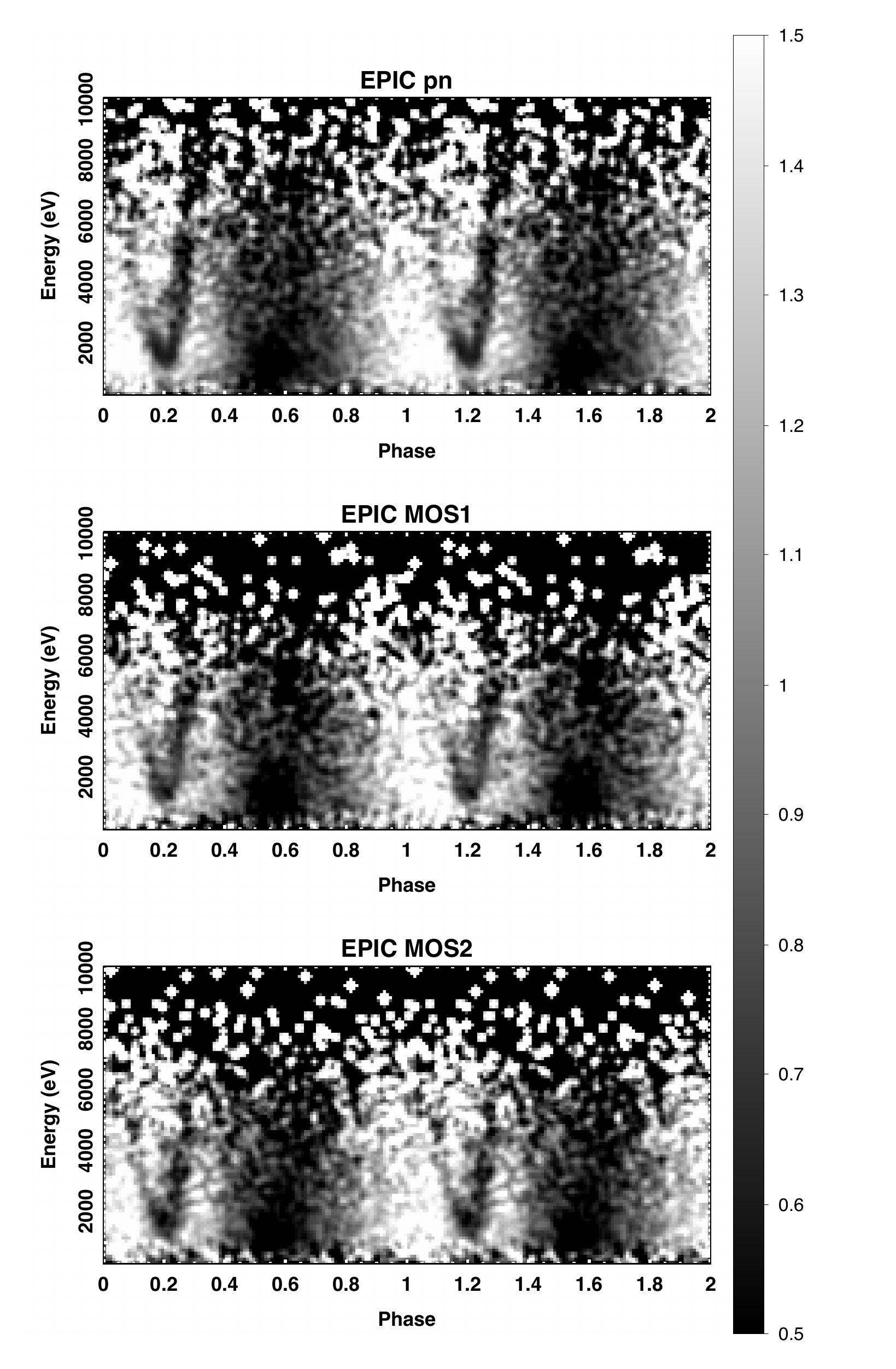}
\caption{Energy versus phase image obtained by binning the
EPIC pn (upper panel), MOS1 (middle panel) and MOS2 (lower panel)
source counts into 100 phase bins and energy channels of 100 eV.
The number of counts in each phase interval has been divided by
the average number of counts in the same energy bin (i.e., the
phase-averaged spectrum).}
\end{figure}

\clearpage

\begin{figure}
\centering\includegraphics[angle=0, height=18cm]{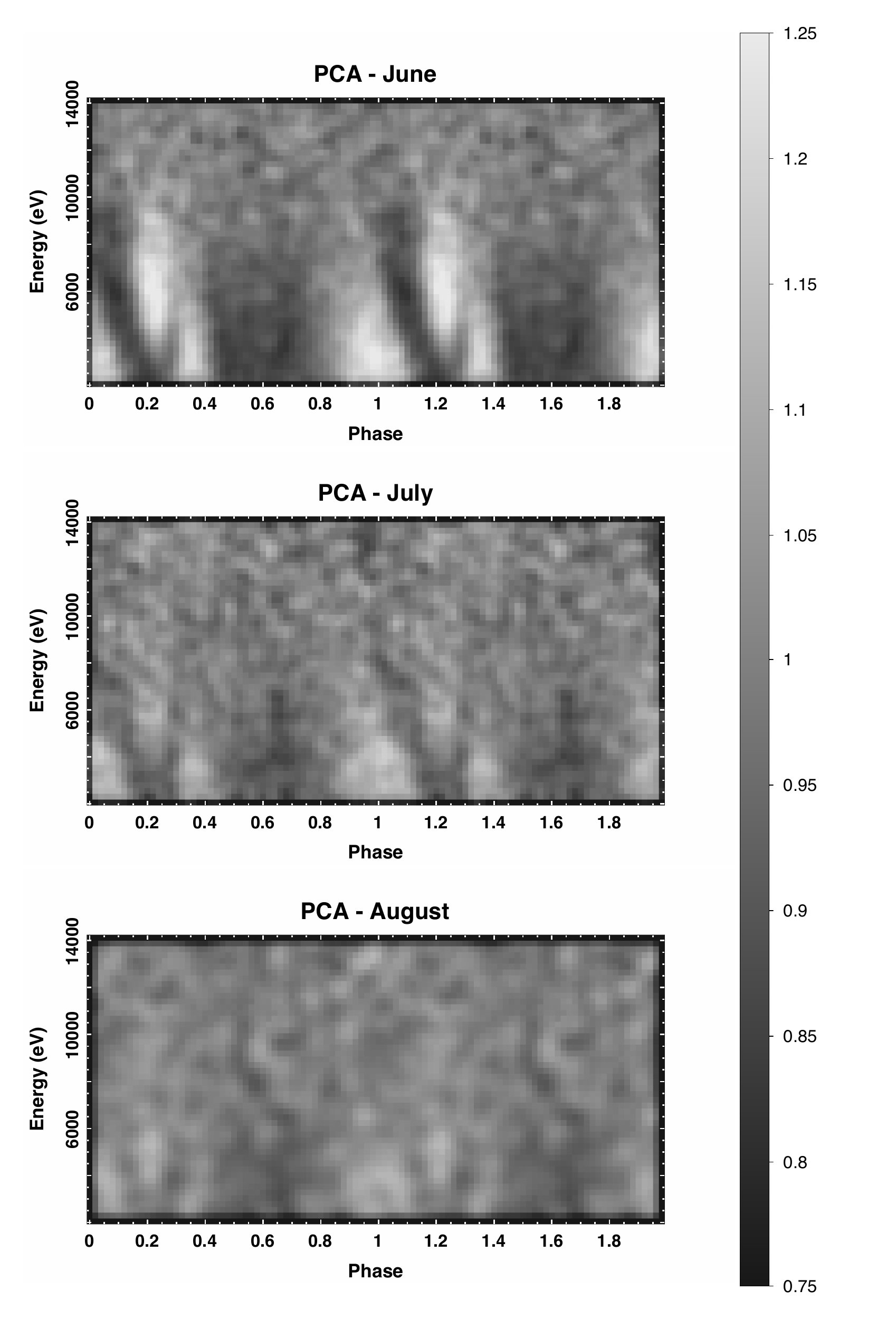}
\caption{
Energy versus phase image obtained by binning the RXTE/PCA data taken in different observations of \src\ (all observations performed in 2009 June, July and August in the upper, middle and lower panel, respectively) into 50 phase bins (two cycles are displayed) and 50 energy channels (channel width 250 eV). The energies within each PCA energy channel have been randomised and the counts in each phase interval divided by the phase-averaged counts in the same energy bin.}
\end{figure}

\clearpage

\begin{figure}
\centering\includegraphics[angle=0, width=15cm]{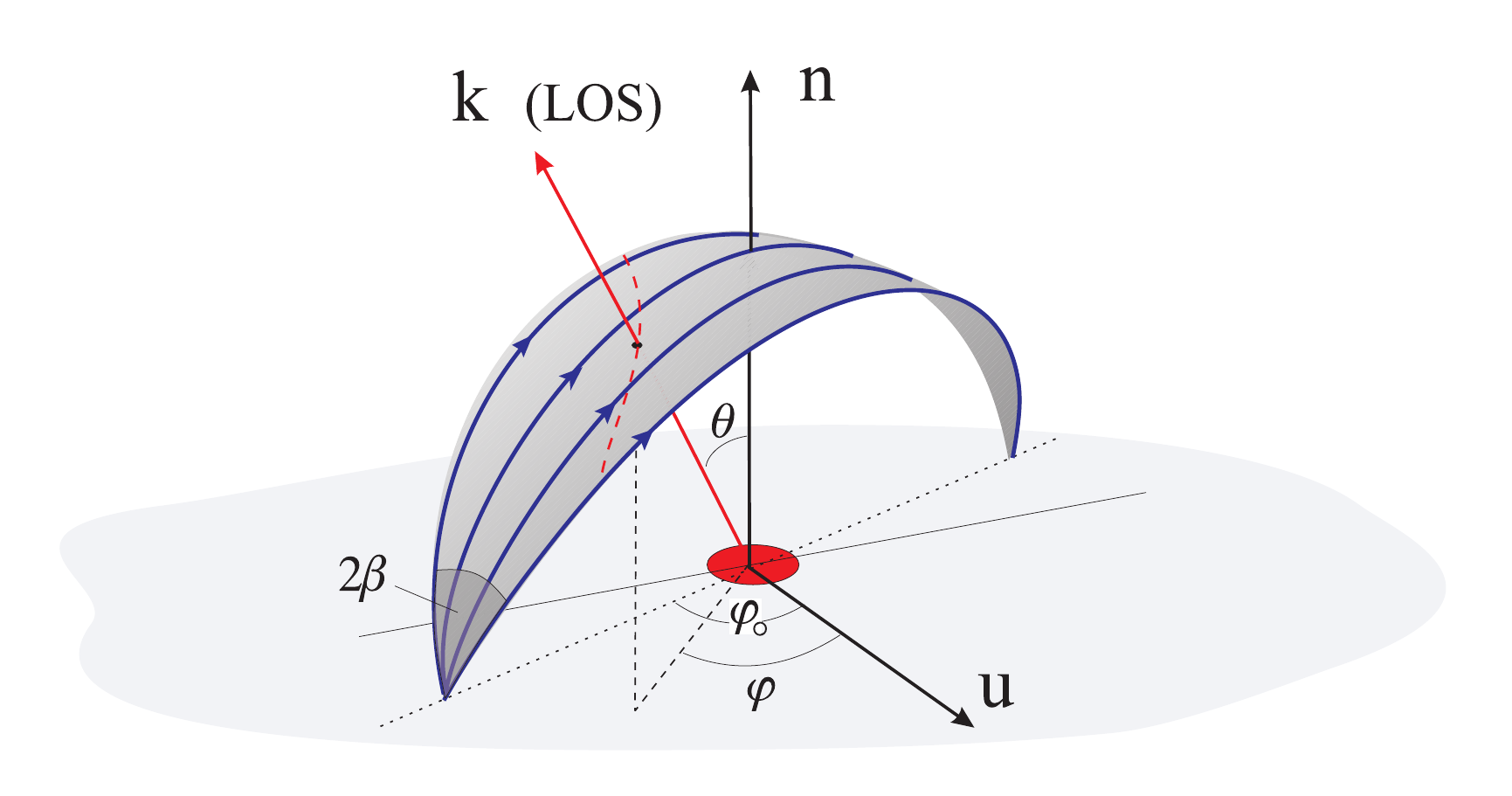}
\caption{ Schematic view of the model discussed in the text.
The line-of-sight (LOS) forms an angle $\theta$ with the normal to
the surface at the spot position, $\bf n$, and crosses a magnetic
loop oriented at some arbitrary direction with respect to $\mathbf u$. General
relativistic ray bending is neglected here, so $\alpha=\theta$.
The actual direction of the photon, $\mathbf k'$, differs from
$\mathbf k$ (see text and equation 3). However, since $\mathbf
k'$, $\mathbf k$ and $\mathbf n$ lie in the same plane, both
directions share the same azimuthal angle $\phi$. }
\end{figure}

\begin{table}
\centering \caption{Parameters of the proton cyclotron model producing the line energy variability displayed in Fig.\,1 (see Fig.\,7 for a schematic view of the model geometry).} 
\vspace{0.5cm}
\label{mod par}
\begin{tabular}{@{}clc}
\hline
Parameter & Description & Value\\
\hline
$\xi$ & Angle between the spot normal and the rotation axis &  $20^{\circ}$\\
$\chi$ & Angle between the LOS and the rotation axis  & $70^{\circ}$ \\
$\phi_0$ & Angle between the lune diameter and the spot meridian  & $90^{\circ}$ \\
$\beta$ & Semi-amplitude of the lune transverse angle  & $30^{\circ}$ \\
$\beta_{\mathrm{c}}$ & Lune inclination angle  & $35^{\circ}$ \\
$z$ & Gravitational redshift at the neutron star surface  & $0.25$ \\
$B_{\mathrm{max}}$ & Magnetic field intensity at the base of the loop  & $7.41\times10^{15}~\mathrm{G}$\\
$f$ & Linear term in Eq.\,6  & $7.16\times10^{15}~\mathrm{G}$\\
\hline
\end{tabular}
\end{table}

\end{document}